\documentclass[letterpaper, 10 pt, conference]{ieeeconf}  

\IEEEoverridecommandlockouts                              
\let\proof\relax

\usepackage{hyperref}       
\usepackage{url}            
\usepackage{booktabs}       
\usepackage{amsfonts}       
\usepackage{nicefrac}       
\usepackage{microtype}      
\usepackage{amsmath,amssymb,amsfonts}
\usepackage{algorithmic}
\usepackage{mathrsfs}
\usepackage{graphicx}
\usepackage{textcomp}
\usepackage{xcolor}
\usepackage{dsfont}
\usepackage{amsthm}
\newtheorem{theorem}{Theorem}[]
\newtheorem{corollary}{Corollary}[]
\newtheorem{lemma}[]{Lemma}

\newtheorem*{assumption*}{Assumption}
\newtheorem{assumption}{Assumption}
\newtheorem{prop}{Proposition}
\newtheorem{definition}{Definition}

\usepackage{amsmath}

\DeclareMathOperator*{\E}{\mathbb{E}}
\DeclareMathOperator*{\R}{\mathbb{R}}

\usepackage{mathtools}
\graphicspath{ {./images/} }
\usepackage{dsfont}
\usepackage[ruled,vlined]{algorithm2e}
\date{}

\title{\LARGE \bf
 Model-Agnostic Zeroth-Order Policy Optimization for Meta-Learning of Ergodic Linear Quadratic Regulators
}

\author{Yunian Pan and Quanyan Zhu$^*$
\thanks{$^*$The authors are with the Department of Electrical and Computer Engineering, Tandon School of Engineering, New York University, Brooklyn, NY, 11201 USA; E-mail: {\tt\small \{yp1170,qz494\}@nyu.edu}}%
}

\begin{document}

\maketitle
\thispagestyle{empty}
\pagestyle{empty}

\begin{abstract}
Meta-learning has been proposed as a promising machine learning topic in recent years, with important applications to image classification, robotics, computer games, and control systems. 
In this paper, we study the problem of using meta-learning to deal with uncertainty and heterogeneity in ergodic linear quadratic regulators. 
We integrate the zeroth-order optimization technique with a typical meta-learning method, proposing an algorithm that omits the estimation of policy Hessian, which applies to tasks of learning a set of heterogeneous but similar linear dynamic systems. 
The induced meta-objective function inherits important properties of the original cost function when the set of linear dynamic systems are meta-learnable, allowing the algorithm to optimize over a learnable landscape without projection onto the feasible set. 
We provide a convergence result for the exact gradient descent process by analyzing the boundedness and smoothness of the gradient for the meta-objective, which justify the proposed algorithm with gradient estimation error being small. We also provide a numerical example to corroborate this perspective.

\end{abstract}

\section{INTRODUCTION}


The Linear Quadratic Regulator (LQR) problems have been thoroughly studied in the setting that all the model parameters are fully known, with its featured Riccati equation being derived using Hamilton-Jacobi equation. 
In the discrete infinite horizon setting, optimal planning can be achieved by solving an Algebraic Riccati Equation (ARE), and computing the optimal control gain accordingly. 
Solutions to the ARE are versatile. One standard method is to iterate the Lyapunov equation. 
There are also other approaches including Semi-Definite Programming (SDP) formulation and methods involving eigenvalue decomposition.
However, these approaches require a significant amount of information of the system parameters, and also significant computation when the sizes of these parameters are large. 
Therefore, there has been a shift to the Reinforcement Learning (RL) paradigm recently to overcome these difficulties, as the complete and precise knowledge of the system parameters is often unavailable. 
In such an incomplete information situation, agents are allowed to interact with either a simulated black-box or the real noisy environment to gain knowledge of the system parameters (the exploration step), while making decisions to optimize the associated performance index (i.e., the exploitation step). 
Typically, RL approaches can be categorized into two classes: the \textit{model-based} approaches and the \textit{model-free} approaches. 

\textit{Model-based} RL starts with the estimation of the unknown system parameters, based on which the controller is constructed. It often follows the certainty equivalence principle: the explicit model parameters are estimated using observations from the system trajectory, the policy constructed is calculated by viewing the estimation as the ground truth. 
See \cite{fiechter1997pac} for the earliest example of offline analysis, and \cite{abbasi2011online}\cite{pmlr-v97-cohen19b}\cite{DBLP:journals/corr/Abbasi-YadkoriS14}\cite{pmlr-v54-abeille17b} for the online counterpart.
\textit{Model-free} RL avoids the process of estimating system parameters. This paradigm has also been developed and can be dated back to \cite{bradtke1994adaptive}, where the authors construct a policy-improvement scheme based on estimated $Q$-functions. 
However, instead of learning a $Q$-function, a more natural approach is to perform policy gradient steps to learn the actual optimal policy gain. This procedure is referred to as policy optimization (PO). As firstly shown by \cite{fazel2018global}, in the setting of deterministic transition dynamics, the policy gradient method converges to the global optimum with polynomial sample complexity. In \cite{DBLP:journals/corr/abs-2011-10300}, the setting of finite horizon and noisy transition dynamics has also been studied. 

RL is useful for solving one specific tasks, but often fails to generalize in heterogeneous and unpredictable environment.
Consider the scenario of learning a set of Linear Time-Invariant systems (LTIs) grouped together by sharing a common state space and control space. Such a group of tasks is usually characterized by heterogeneous dynamics, uncertainties, and control purposes. E.g., a robot arm may need to adapt its control strategy to different human movements such as grabbing, lifting, or pushing; or an industrial power plant is often in need to deal with different load demands and external disturbance. 
While classical $H_{\infty}$-methods take a pessimistic approach to counter the capacity of external disturbance by introducing a zero-sum game, 
the recent Model-Agnostic Meta-Learning (MAML) approach \cite{finn2017model} appears to be more promising in dealing with model-heterogeneity in benign environment. 
The concept of meta-learning arose from the idea that intelligent agents should be adaptable to different learning tasks. 
Many meta-learning algorithms have been proposed in recent years, and MAML is one typical method among them.
MAML learns a parameter that is tunable to a variety of optimization problems in the sense that whenever the agent encounters a new task, a few gradient updates on the parameter will attain good performance. 

This motivates us to explore the connection between meta-learning-based RL and LQR. We focus on integrating MAML with the model-free PO methods, since learning the shared structure of the model is less efficient and explicit than learning a shared initial policy. 
We propose an algorithm to refine an optimal initial policy parameter for a set of ``similar'' LTIs. 
The algorithm relies on zeroth-order optimization techniques as a first order oracle is not available. 
We will dive into the properties of the  meta-gradient, which establish the convergence of exact gradient gradient process, and hence the actual algorithmic convergence given the estimation error being small. 
The essential step is to bound the operator norm of policy Hessian for original cost function, henceforth the Frobenius norm of the meta-gradient can be bounded as well. 
Overall, unlike typical MAML methods, our method overcomes the barrier of estimating second order information in constructing the meta-gradient by directly exploiting the perturbation trick. 
A numerical experiment will be demonstrated as an empirical validation for the algorithm.

\section{RELATED WORK}
PO methods can be dated back to the model-based approach in 1970 \cite{gershwin1970discrete} when it was called differential dynamical programming. Since modeling details are not available in practice, there has been a variety of estimation methods emerging, including finite-difference methods and REINFORCE \cite{williams1992simple}. Our framework is built upon finite-difference methods but estimating the gradient in a different fashion \cite{fazel2018global}\cite{DBLP:journals/corr/abs-2011-10300}. We extend such technique to meta-learning case, and develop a new meta-gradient estimation procedure. 

MAML is proposed by \cite{finn2017model} as an optimization-based meta-learning method. Its convergence was analyzed in the framework of finite-horizon Markov decision processes by \cite{fallah2020provably}. Followed by \cite{molybog2020global}, which studies the global convergence of MAML over LQR tasks. However, their results rely on the global property assumption of the cost function and the results are generalized to any cost function that has such property, while our work gives more details about the set of specific LTIs and how the learning proceeds.
Unlike the formal framework in \cite{fallah2020provably}, our procedure avoids estimating the second order information, hence eliminating the potential errors arising from the policy Hessian estimation.

\section{PROBLEM FORMULATION}

\subsection{Policy Optimization for LQR}

Let $\mathcal{S} = \{(A_i, B_i, Q_i, R_i, \Psi_i)\}_i$ be the set of LQR systems, where $A_i \in \R^{ d \times d}, B_i \in \R^{d \times k}$ are system dynamics matrices, $Q_i \in \R^{ d \times d} , R_i \in \R^{k \times k}, Q_i, R_i \succeq 0$ are the associated cost matrices, and $\Psi_i \in \R^{d \times d}$ are the noise covariance matrices, which are symmetric and positive definite. We can sample the black-box systems from a probability distribution $p$. Each system $i$ is assumed to share the same state space $ \R^d$ and control space $ \R^k$, and is governed by the linear system dynamics associated with the quadratic cost functions: 
\begin{equation*}
    x_{t+1} = A_i x_{t} + B_i u_{t} + w_t, \quad\quad g_i(x_t, u_t) = x_t^{\top}Q_i x_t + u_t^{\top} R_i u_t ,
\end{equation*}
where  $x_t \in \R^d$, $u_t \in \R^k$, $w_t \sim \mathcal{N} (0 , \Psi_i)$ is the random i.i.d. noise with covariance matrix $\Psi_i$. 

For each system $i$, our objective is to minimize the average infinite horizon cost, 
\begin{equation*}
    J_i = \lim_{T \to \infty}\frac{1}{T}\mathbb{E}_{x_0 \sim \rho_0, \{w_t\} }\left[\sum_{t-0}^{T-1}g_i(x_t, u_t)\right],
\end{equation*}
where $\rho_0$ is the initial state distribution $\mathcal{N}(0, \Sigma_0)$. 
It's well-known that the optimal policy is of linear form in state, i.e., for system $\mathcal{S}_i$, the optimal control $\{u_t^{i*}\}_{t \geq 0} $ can be expressed as $u^{i*}_t = - K^{i*} x_t$, where $K^{i*} \in \R^{k \times d}$ satisfies $K^{i*}=\left(R_i+B_i^{\top} P_i^{*} B_i\right)^{-1} B_i^{\top} P^{i*} A_i$, and $P^{i}_*$ is the unique solution to the following discrete algebraic Riccati equation 
$ P^{i}_*=Q_i+A_i^{\top} P^{i}_* A_i+A_i^{\top} P^{i}_*  B_i\left(R_i+B_i^{\top} P^{i}_* B_i\right)^{-1}\ B_i^{\top} P^{i}_* A_i$.

A policy $K \in \R$ is called stable for system $i$ if.f $\rho(A_i - B_i K) < 1$, where $\rho(\cdot)$ stands for the spectrum radius of a matrix.
Denoted by $ \mathcal{K}_i$ the set of stable policy for system $i$.
For a policy $K \in \mathcal{K}_i$, the induced cost over system $i$ is 
\begin{equation*}
\begin{aligned}
     J_i(K) & = \lim_{T \to \infty}\frac{1}{T}\mathbb{E}_{x_0 \sim \rho_0, w_t }[\sum_{t-0}^{T-1}\left(x_{t}^{\top}( Q_i + K^{\top}R_iK) x_{t}\right)] \\
      = \E_{x \sim \rho^i_K} &[x^{\top}( Q_i + K^{\top}R_iK) x]  = \operatorname{Tr}\left[(Q_i + K^{\top} R_i K) \Sigma^i_K \right],
\end{aligned}
\end{equation*}
where the limiting stationary distribution of $x_t$ is denoted by $\rho^i_K$, $\operatorname{Tr}(\cdot)$ stands for the trace operator. The Gramian matrix $\Sigma^i_K:= \mathbb{E}_{x \sim \rho_K^i} [xx^{\top}] = \lim_{T \to \infty} \E_{x_0 \sim \rho_0 } [\frac{1}{T} \sum_{t=0}^{T-1} x_t x_t^{\top}]$ satisfies the following Lyapunov equation
\begin{equation}\label{xgramian}
    \Sigma^i_{K}=\Psi_i +(A_i-B_i K) \Sigma^i_{K}(A_i-B_i K)^{\top}.
\end{equation}
\eqref{xgramian} can be easily verified through elementary algebraic manipulation. 

The following lemma gives an explicit form of the policy gradient of a single LQR.

\begin{prop}[Policy Gradient of Ergodic LQR \cite{DBLP:journals/corr/abs-1907-06246}] \label{prop1}
The expression for average cost is $J_i(K) = \operatorname{Tr}(P^i_K \Psi_i)$, 
and the expression of $\nabla J_i(K)$ is 
 \begin{equation}
   \begin{aligned}
       \nabla J_i(K)  &= 2\left[\left(R_i+B_i^{\top} P_{K} B_i\right) K-B_i^{\top} P^i_{K} A_i\right] \Sigma^i_{K}\\ &=2 E^i_{K} \Sigma^i_{K}
   \end{aligned}
 \end{equation}
 where $\Sigma^i_{K}$ satisfies \eqref{xgramian}, $E^i_K$ is defined to be
 \begin{equation*}
     E^i_K := \left(R_i+B_i^{\top} P^i_{K} B_i\right) K-B_i^{\top} P^i_{K} A_i,
 \end{equation*}
 and $P^i_K$ is the unique positive definite solution to the Bellman equation.
 \begin{equation*}
     P^i_{K}=\left(Q_i+K^{\top} R_i K\right)+(A_i-B_i K)^{\top} P^i_{K}(A_i - B_i K).
 \end{equation*}
\end{prop}

Note that this result itself is insufficient for practice since the system parameters are unknown to us. 
However, it is indispensable for us to explore the coercivity, almost smoothness, and gradient dominance property of the cost function, providing solid theoretical evidence for policy optimization methods.



\subsection{Zero-th Order Method for Gradient-Estimation}
In practice, it is usually challenging to evaluate both $E^i_K$ and $\Sigma^i_K$ due to the absence of knowledge about system matrices. 
This can be done approximately by evaluating $K$ and collecting trajectories from the system. 
Denote the dataset obtained using $K$ as $\mathcal{D}_K^i$. In practice, 
it can be a set of trajectories obtained solely using $K$, or trajectories obtained from policy perturbed around $K$. 
The latter is usually coupled with the zeroth-order methods. 

The zeroth-order methods are derivative-free optimization techniques that allow us to optimize an unknown smooth function $J_i(\cdot): \R^{k \times d} \to \R$ by estimating the first-order information. What it requires is to query the function values $J_i$ at some input points. 
A generic procedure is to firstly sample some perturbations $U \sim \operatorname{Unif}(\mathbb{S}_r)$, where $\mathbb{S}_r$ is a $r$-radius $k \times d$-dimensional sphere, and estimate the gradient of the perturbed function through equation:
\begin{equation}\label{steinidtty}
    \nabla_r J_i(K) = \frac{ dk }{ r^2} \E_{U \sim \operatorname{Unif}(\mathbb{S}_r)}[ J_i(K+U)U].
\end{equation}
The expectation can be evaluated through Monte-Carlo methods. \
During the learning of LQR systems, there does not necessarily exist a function value oracle, i.e., the value of $J_i$ is not accessible. 
One simple approach is to sample multiple trajectories to estimate the return. 
Also, the perturbed variable $K+U$ might be inadmissible. This can be resolved by restricting $r$ to be small such that the change on $K$ is not drastic.
Algorithm \ref{zerothorderalgo} (adapted from \cite{fazel2018global}) gives an example of gradient-estimation procedure.

\begin{algorithm}[htbp]
\label{zerothorderalgo}
\SetKwInOut{Input}{Input}
\SetAlgoLined
\Input{Simulator $i$, Policy $K$, number of trajectories $M$, roll out length $\ell$, parameter $r$}
\For{$m = 1, 2, \ldots, M$}{
     Sample a policy $\widehat{K}_m = K + U_m$, where $U_m$ is drawn uniformly over random matrices whose Frobenius norm is $r$ \;
     Simulate $\widehat{K}_i$ for $\ell$ steps starting from $x_0 \sim \rho_0$. Let $\widehat{J}_m$ and $\widehat{\Sigma}_m$ be empirical estimates
     $$
         \widehat{J}_{m}= \frac{1}{\ell}\sum_{t=1}^{\ell} g_{t}, 
     $$
     where $g_t$ and $x_t$ are costs and states of the current trajectory $m$.
}
Return the (biased) estimates:
\begin{equation*}
     \widehat{\nabla} J_i(K) = \frac{1}{M} \sum_{i=1}^M \frac{dk}{ r^2} \widehat{J}_m U_m,  
 \end{equation*}
\caption{\texttt{Gradient Estimation} \cite{fazel2018global}}
\end{algorithm}

Algorithm \ref{zerothorderalgo} enables us to perform iterations similar to $K^{\prime} = K - \eta \widehat{\nabla}J_i(K)$. 
However, since this optimization has inexplicit constraints that $K^{\prime}$ must be stable, it is questionable how small $r$ should be to prevent $K^{\prime}$ from escaping the admissible set. 
Lemma \ref{boundlipl} quantifies how small one-step update should be in order for the updated policy to be still stable.

\begin{lemma}[Perturbation analysis adapted from \cite{fazel2018global}]
\label{boundlipl} 
 For any $i \in \mathcal{S}$, suppose $K^{\prime}$ is small enough such that:
 \begin{equation*}
     \|K - K^{\prime} \| \leq \min \left( \frac{\sigma_{\min}(Q_i) \sigma_{\min}(\Psi_i)}{4 J_i(K) \|B_i \| (\|A_i - B_i K\|+ 1)}, \|K\|\right),
 \end{equation*}
 then there is a polynomial $L_J$ in $\frac{J_i(K)}{\sigma_{\min}(\Psi_i) \sigma_{\min}(Q_i)}$, $\sigma_{\min}(\Psi_i)$, $\frac{1}{\sigma_{\min}(R_i)}$, $\| A_i \|$, $\| B_i\|$, $\|R_i\|$ such that 
 \begin{equation*}
     \| \nabla J_i (K) - \nabla J_i (K^{\prime})\|_F \leq L_J \| K -K^{\prime}\|_F.
 \end{equation*}
\end{lemma}

\subsection{The Meta-Learning Problem}
In analogy to \cite{finn2017model,fallah2020provably}, the meta-learning problem can be formulated as finding a meta-policy initialization, such that one step of stochastic policy gradient step still attains good performance for systems $\mathcal{S}$ in expectation:
\begin{equation} \label{obj}
    \min_{K \in \bigcap_i \mathcal{K}_i} \mathcal{L}(K) := 
\mathbb{E}_{i \sim p}\left[\mathbb{E}_{\mathcal{D}_K^i}J_{i}\left( K  - \eta \widehat{\nabla} J_{i}\left(K, \mathcal{D}_{K}^{i}\right)\right)\right] ,
\end{equation}
where $\eta$ is the learning rate, $\widehat{\nabla} J_{i}\left(K, \mathcal{D}_{K}^{i}\right)$ is the gradient estimation obtained from dataset $\mathcal{D}_K^i$.

To analyze policy optimization for the meta-learning problem, we begin with analyzing the gradient of function \eqref{obj}, given by a vectorized version:
\begin{equation}\label{derivative}
\begin{aligned}
         \operatorname{vec}(\nabla \mathcal{L}&(K))  = \operatorname{vec}(\nabla \mathbb{E}_{i \sim p, \mathcal{D}_K^i}J_{i}\left( K  - \eta \widehat{\nabla} J_{i}\left(K, \mathcal{D}_{K}^{i}\right)\right)  )   \\
         & = \mathbb{E}_{i \sim p, \mathcal{D}_K^i}\bigg[(I - \eta \nabla (\operatorname{vec}(\widehat{\nabla} J_i (K, \mathcal{D}_K^i))))  \\ & \quad\quad \operatorname{vec}(\nabla J_i(K^{\prime}))\bigg]        +  \mathbb{E}_{i \sim p} \nabla \mathbb{E}_{\mathcal{D}_K^i} \left[J_i(K^{\prime})\right]         \\
         & = \mathbb{E}_{i \sim p, \mathcal{D}_K^i}\bigg[ (I - \eta \operatorname{mat}(\nabla^2 J_i (K)) ) \operatorname{vec}(J_i(K^{\prime})) \bigg]  \\ 
         & \quad + \mathbb{E}_{i \sim p} \int_{\mathcal{D}_K^i} J_i(K - \eta \widehat{\nabla}J_i(K, \mathcal{D}_K^i)) \nabla d\mathbb{P}(\mathcal{D}_K^i),
\end{aligned}
\end{equation}
where $K^{\prime} = K - \eta \widehat{\nabla} J_i(K,\mathcal{D}_K^i )$, $\operatorname{vec}$ is the vectorization of a matrix and $\operatorname{mat}$ is the matrices reshaping of a 4d-tensor, $d\mathbb{P}(\cdot)$ is the measure for obtaining a particular set of trajectories $\mathcal{D}_K^i$ by performing $K$. 
Here, first we argue that $\nabla (\widehat{\nabla})$ can be replaced by $\widehat{\nabla}^2$ or $\nabla^2$ when an empirical estimate of $\nabla \mathcal{L}(K)$ is close to itself.
Second, since we adopt a deterministic policy $K$, the trajectory sampling randomness only depends on the randomness of $U_m$, $m= 1, \ldots, M$, which is independent of $K$, the second term in \eqref{derivative} becomes $0$. 
Thus, we avoid recursively writing $\mathcal{D}_K^i$ and arrive at a simple formula:
\begin{equation}
\label{refinedgrad}
   \operatorname{vec} (\nabla \mathcal{L}(K)) = \mathbb{E}_{i \sim p} (I - \eta \operatorname{mat} (\nabla^2 J_i (K )))  \operatorname{vec}(\nabla J_i(K^{\prime})) .
\end{equation}

It turns out estimating the terms $\nabla^2 J_i (K)$ and $\nabla J_i (K)$ separately is still hard. We take a different route that directly exploits the zeroth-order method to estimate $\nabla \mathcal{L}(K) $.

\subsection{Zero-th Order Method for Meta-Gradient Estimation}

Recall \eqref{steinidtty} and its practical variant for function $\mathcal{L}$, we have the gradient expression for perturbed $\mathcal{L}$:
\begin{equation*}
    \nabla_r \mathcal{L} (K) = \frac{dk}{r^2} \E_{U \sim \mathbb{S}_r, i \sim p} \left[J_i (K + U - \eta \nabla J_i (K+ U) ) U\right] .
\end{equation*}
To evaluate expectation $\E_{U \sim \mathbb{S}_r, i \sim p}$ we sample $D$ independent perturbation $U_d$ and a batch of systems $\mathcal{S}_n$, then average the samples. 
To evaluate return $J_i(K+U - \eta \nabla J_i(K+U))$ we first apply algorithm \ref{zerothorderalgo} to obtain approximate gradient for a single perturbed policy, then roll out the one-step updated perturbed policy to estimate its associated return. 

A comprehensive description of the procedure is shown in Algorithm \ref{zerothordermethods}. Essentially we aim to collect $D$ samples for return by perturbed policy $\hat{K}_d^i$, which requires the original perturbed policy $\hat{K}_d$ and the gradient estimate of it. To do so, we use Algorithm \ref{zerothorderalgo} as an inner loop procedure. After computing $\hat{K}_d^i$ we simulate it for $\ell$ steps to get the empirical estimate of return $J_i(K+U - \eta \nabla J_i (K+U))$.

\begin{algorithm}[htbp]
\label{zerothordermethods}
\SetKwInOut{Input}{Input}
\SetAlgoLined
\Input{Meta-environment $p$, policy $K$, number of meta-perturbations $D$,  inner-perturbations $M$ learning rate $\eta$, roll-out length $\ell$, parameter $r$; }
Randomly draw systems batch $\mathcal{S}_n $ from meta-environment $p$ \;
\For{all $i \in \mathcal{S}_n$}{
     \For{$d = 1, 2, \ldots, D$}{
      Sample a policy $\widehat{K}_d = K + U_d$, where $U_d$ is drawn uniformly over random matrices whose Frobenius norm is $r$ \;
      Estimate $\widehat{\nabla}J_i(\widehat{K}_d) \leftarrow \texttt{Gradient Estimation}(i, \widehat{K}_d, \ell, r)$ \;
      Perform one step gradient descent:
      \begin{equation} \label{onestepinalgo}
          \widehat{K}_d^i = \widehat{K}_d - \eta \widehat{\nabla} J_i(\widehat{K}_d) ;
      \end{equation}
      
      Estimate $\hat{J}_{i}(K^i_d)$ from simulating $K^i_d$ for $\ell$ steps starting with $x_0 \sim \rho_0$:
      $$
       \hat{J}_{i,d}(K^i_d) = \frac{1}{\ell} \sum_{t=1}^{\ell} g_t ,
      $$
      where $g_t$ are the costs per time step.
     }
}
The meta-gradient is:
     \begin{equation*} 
         \widehat{\nabla} \mathcal{L}(K) = \frac{1}{|\mathcal{S}_n|}\sum_{i \in \mathcal{S}_n} \frac{1}{D} \sum_{d = 1}^D \frac{dk}{r^2} \widehat{J}_{i}(K^i_d) U_d
     \end{equation*}
\caption{\texttt{Meta-Gradient Estimation}}
\end{algorithm}
It is worth noting that even though the return estimate is biased, intuitively we expect it to be close to the true return if the roll-out length $\ell$ is large enough. Theoretically quantifying such distance will be research of future interests.
In summary, the meta-learning algorithm is shown in algorithm \ref{metalqr}.

\begin{algorithm}[htbp]
\label{metalqr}
\SetKwInOut{Input}{Input}
\SetAlgoLined
\Input{ Meta-environment $p$, number of meta-perturbations $D$, inner-perturbations $M$, learning rate $\eta$, learning rate $\alpha$, roll-out length $\ell$, parameter $r$, tolerance $\varepsilon$;}
Initialize feasible policy $K \in \bigcap_{i \in \mathcal{S}} \mathcal{K}_i$ \; 
\While{$ \| \widehat{\nabla}\mathcal{L}(K) \leq \varepsilon \|$}{
    $\widehat{\nabla} \mathcal{L}(K) \leftarrow $ \texttt{Meta-Gradient Estimation}$(p, K_{n}, D, M, \eta, \ell, r)$ \;
    One step gradient:
    \begin{equation*}
           K_{n+1}= \leftarrow K_{n} - \alpha \widehat{\nabla} \mathcal{L}(K_n);
    \end{equation*}
    
    $n \leftarrow n + 1$
}
\caption{Model-Agnostic Meta Learning for Linear Quadratic Gaussian}
\end{algorithm}

\section{GRADIENT DESCENT ANALYSIS}
It remains to provide a sanity check for the algorithm.
Essentially, the purpose of such an optimization algorithm is to find a first-order stationary point.

\begin{definition}[First-order Stationary Point]
 A control gain $K \in \R^{d \times k}$ is said to be a (first-order) stationary point if and only if $\|\nabla \mathcal{L}(K) \|_F = 0$; similarly, it is said to be an $\varepsilon$-stationary point if and only if $ \| \nabla \mathcal{L} (K)\|_F  \leq \varepsilon$.
\end{definition}
The operator $\|\cdot\|_F$ is the Frobenius norm.

In general, an arbitrary collection of LQRs is not necessarily meta-learnable using gradient-based optimization techniques, as one might not be able to find an admissible initialization of policy gain.

For instance, consider a two-system scalar case where $A_1 = 3, B_1 = 4$ and $A_2 = 1, B_2 = -1$. The policy evaluation requires an initialization $K$ to be stable for both system, which means $K \in (\frac{1}{2}, 1) \cap (-2, 0) = \emptyset$! This example illustrates that in regards to LQR cases, not all collections of LTIs are meta-learnable using MAML.

Thus, it is reasonable to assume that the systems are close to each other in the sense that the intersection of their feasible policy sets admits at least one stationary point. 

\begin{assumption}\label{admissible}
 The following statements are true:
 \begin{itemize}
     \item[1.] The set $\bigcap_{i \in \mathcal{S}} \mathcal{K}_i \neq \emptyset$. 
     \item[2.] The set $\bigcap_{i \in \mathcal{S}} \mathcal{K}_i$ contains at least one stationary point. 
 \end{itemize}
\end{assumption}

Recall that the algorithm produces a close approximation for iteration:
\begin{equation}
\label{thegoaliteration}
    K_{n+1} = K_n -  \frac{\alpha}{|\mathcal{S}_n|} \sum_{i }(I - \eta \operatorname{mat}\nabla^2 J_i(K)) \operatorname{vec}\nabla J_i(K^{\prime}),
\end{equation}
it suffices to show \eqref{thegoaliteration} converges to a stationary point.
In order to analyze such an exact gradient descent process, we need to quantify the term $\nabla^2 J_i(K)$, which is actually a 4-dimensional tensor. To avoid using tensor algebra, we consider scalar case as a simplified version of proof. The reason is that often in analyzing an unbiased stochastic gradient algorithm, it suffices to show the boundedness and smoothness of gradient $\nabla \mathcal{L}$ in the sense of its Frobenius norm. In particular, smoothness almost implies convergence to first order stationary point.

\subsection{On the boundedness and Lipschitz Property of $\nabla \mathcal{L}(K)$}

In the following, we characterize boundedness and Lipschitz property of $\nabla \mathcal{L}(K)$. With a slight abuse of notations, we let $\|\nabla^2 J_i(K)\|$ be the operator norm of $\operatorname{mat}(\nabla^2 J_i(K))$, which can be bounded by lemma \ref{boundhessianj}.
\begin{lemma}\label{boundhessianj}
The operator norm of the Hessian $\nabla^2 J_i(K)$ is bounded by the following:
\begin{equation*}
    \| \nabla^2 J_i(K)\|  \leq 2c^i_1  + 4c^i_2
\end{equation*}
where 
\begin{equation*}
\begin{aligned}
       c^i_1 & := \left( \| R_i\| + \| B_i \|^2 \frac{J_i(K)}{\sigma_{\min}(\Psi_i)} \right)\cdot \frac{J_i(K)}{\sigma_{\min}(Q_i)} \\
       c^i_2 & := c^i_3 \| A_i - B_i K\| \| B_i\| \frac{J_i(K)}{\sigma_{\min}(Q_i) \sigma_{\min}(\Psi_i)} \\
       c^i_3 & := \frac{J_i(K)}{\sigma_{\min}(Q_i)}\left( \| R_i \| \|K\| + \| B_i\| \|A_i - B_iK\| \frac{J_i(K)}{\sigma_{\min}(\Psi_i)} \right)
\end{aligned}
\end{equation*}
\end{lemma}

Leveraging \ref{boundhessianj}, we arrive at lemma \ref{boundgradientl} that bounds the Frobenius norm of $\nabla \mathcal{L}(K)$.
\begin{lemma}\label{boundgradientl}
The Frobenius norm of the gradient $\| \nabla \mathcal{L}\|_F$ is bounded by:
\begin{equation}
\begin{aligned}
       \| \nabla \mathcal{L}(K)\|_F & \leq \E_{i \sim p} [(\| I \| +  \eta \| \nabla^2 J_i(K) \| ) \|\nabla J_i(K^{\prime})\|_F ]\\
       & \leq \max_{i \in \mathcal{S}} (1 + \eta (2c^i_1 + 4c^i_2)) \cdot 2c^i_4
\end{aligned}
\end{equation}
where $c^i_1$ and $c^i_2$ are defined in lemma \ref{boundhessianj}, 
$$
c^i_4  := \frac{ J_i(K)d}{\sigma_{\min}(Q_i)}( \| R_i K\|_F + d\| B_i(A_i - B_iK)\|_F \frac{ J_i(K)d }{\sigma_{\min}(\Psi_i)} ).
$$

\end{lemma}

Equipped with the results above, we are able to characterize that there exists a polynomial bound on the Frobenius norm of the gradient. 

\begin{corollary}
 For any $K \in \bigcap_{i \in \mathcal{S}}\mathcal{K}_i$, there exists a polynomial $G$ in  $\frac{J_i(K)}{\sigma_{\min}(\Psi_i) \sigma_{\min}(Q_i)}$, $\sigma_{\min}(\Psi_i)$, $\frac{1}{\sigma_{\min}(R_i)}$, $\| A_i \|$, $\| B_i\|$, $\|R_i\|$, and $d$ such that
 \begin{equation*}
     \| \nabla \mathcal{L}(K) \|_F \leq G.
 \end{equation*}
\end{corollary}

Then, we aim to extend lemma \ref{boundlipl} to our meta-learning case. We characterize that if the gradient update is small enough for any system $i$, the updated policy is still inside $\bigcap_i \mathcal{K}_i$. 

\begin{lemma}[Perturbation analysis of $\nabla \mathcal{L}(K)$]
 Suppose $K^{\prime}$ is small enough such that $\forall i \in \mathcal{S}$:
 \begin{equation*}
     \|K - K^{\prime} \| \leq \min \left( \frac{\sigma_{\min}(Q_i) \sigma_{\min}(\Psi_i)}{4 J_i(K) \|B_i \| (\|A_i - B_i K\|+ 1)}, \|K\|\right),
 \end{equation*}
 then there exists a polynomial $L$ in $\frac{J_i(K)}{\sigma_{\min}(\Psi_i) \sigma_{\min}(Q_i)}$, $\sigma_{\min}(\Psi_i)$, $\frac{1}{\sigma_{\min}(R_i)}$, $\| A_i \|$, $\| B_i\|$, $\|R_i\|$ such that 
 \begin{equation*}
     \| \nabla \mathcal{L} (K) - \nabla \mathcal{L} (K^{\prime})\|_F \leq L \| K -K^{\prime}\|_F
 \end{equation*}
\end{lemma}

\proof[Proof Sketch]{
The existence of $L$ is immediate after lemma \ref{boundhessianj} and \ref{boundlipl} since $\nabla \mathcal{L}$ still consists of terms in $\nabla J_i (K)$ and $\nabla^2 J_i (K)$,  the condition is $\|K- K^{\prime}\|$ is small for any $i$.  We omit computing explicit form of $L$ here.
\qedhere
}

\subsection{Convergence of Exact Gradient Descent }

\begin{theorem}
    Suppose the parameter  $\alpha \in (0, 1/L]$, where $L$ is defined in lemma \ref{boundgradientl}. 
    Consider the exact gradient descent process, $K_{n+1} = K_n - \alpha \nabla \mathcal{L}(K_n)$.
    Then, $K_n$ converges to a stationary point, i.e., $\lim_{n \to \infty}\| \nabla \mathcal{L}(K_n) \|_F= 0$. 
\end{theorem}

\proof[Proof Sketch]{
We start by using smoothness property:
\begin{equation}\label{recur}
\begin{aligned}
      | \mathcal{L}(K_{n+1}) & - \mathcal{L}(K_n) -\| \nabla \mathcal{L} (K_n) (K_{n+1} - K_n) \|_F |  \\ & \leq \frac{L^2}{2} \| K_{n+1} - K_n \|^2_F.
\end{aligned}
\end{equation}
Recall at iteration $n+1$
$
    K_{n+1} = K_n - \alpha \nabla \mathcal{L}(K_{n}) ,
$
and plug it into \eqref{recur}, we obtain
\begin{equation*}
\begin{aligned}
     \mathcal{L}(K_{n+1})& \leq \mathcal{L}(K_n) -  \alpha  \| \nabla\mathcal{L}(K_n) \|^2_F  + \frac{L^2}{2}\alpha^2 \|\nabla \mathcal{L}(K_n)\|^2_F \\
     & \leq \mathcal{L}(K_n) - \frac{\alpha}{2} \| \nabla \mathcal{L}(K) \|_F^2
\end{aligned}
\end{equation*}
Above implies that $\mathcal{L}(K_n)$ is monotonically non-increasing, thus either $\mathcal{L} (K_n) \to -\infty$ or else converges to some finite value, since $\mathcal{L}(K_n) > 0$, we have $\| \nabla \mathcal{L}(K_n)\|_F \to 0$. \qedhere
}

Hence we provide a local convergence result for the case of constant stepsize, which follows immediately after the gradient smoothness. 
Since our meta-gradient estimator is generally not unbiased, the actual sample complexity needs further theoretical investigation.
as there are cases where biased stochastic gradient still has convergence guarantee.

\section{Numerical Results}
We consider three cases of state and control dimensions in the numerical example, but due to computational limits, we consider a moderate system collection size $|\mathcal{S}| = 5$.
The collection of systems is randomly generated to behave ``similarly''. 
Specifically, we sample matrices $A_0, B_0, Q_0, R_0, \Psi_0$ from uniform distributions, and adjust $A_0$ so that $\rho(A_0) < 1$, adjust $Q_0, R_0, \Psi_0$ to be symmetric and positive definite. 
Then, we sample the rest of systems $i$ independently such that their system matrices are centered around $A_0, B_0, Q_0, R_0, \Psi_0$, (for example $[A_i]_{m,n} \sim \mathcal{N}([A_0]_{m,n}, 0.25)$ for some $i$, $m$ and $n$.) and follow the same procedure to make $\rho(A_i) < 1$ and $Q_i, R_i, \Psi_i$ positive definite.

\begin{figure}[htbp]
    \centering
    \includegraphics[width = .4\textwidth]{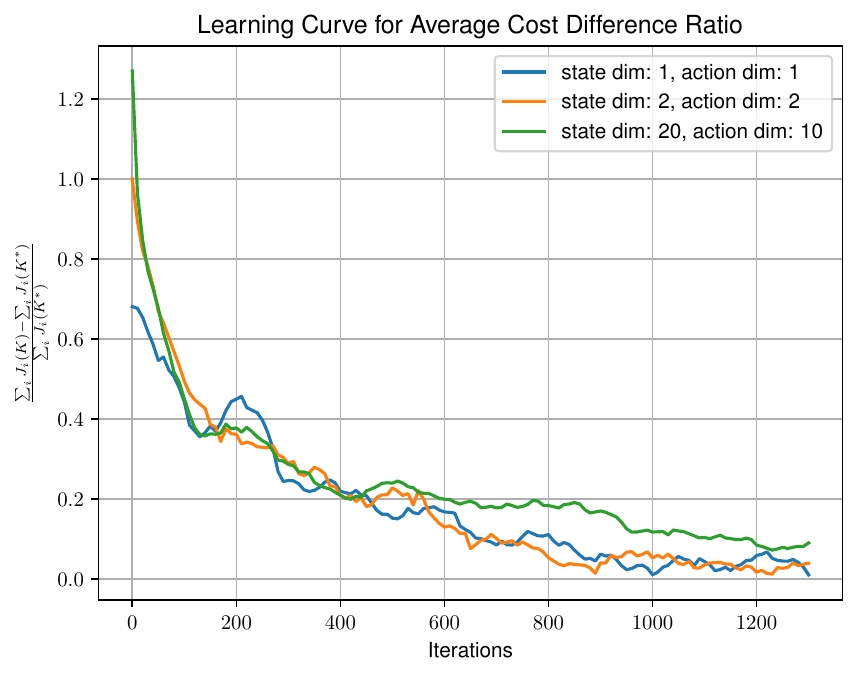}
    \caption{The plot shows three curves encapsulating the changing of average performance during gradient descent, each corresponds to a particular dimension setting of state and action space, (green: $d = 20, k = 10$, orange: $d = 2, k = 2$, blue: $d= 1, k =1$.) constant learning rates $\alpha = 1e-3$, $\eta = 1e-5$ for orange and blue cases and $\alpha = 1e-5$, $\eta = 1e-7$ for green curve, numbers of meta and inner perturbation $D = 100, M= 100$, gradient smooth parameter $r = 0.05$, roll out length $\ell = 50$.}
    \label{plot}
\end{figure}

We report the learning curves for average cost difference ratio $\frac{\sum_{i \in \mathcal{S}} J_i(K_n) - J_i (K^{i*})}{\sum_{i \in \mathcal{S}} J_i(K^{i*}) }$, this quantity captures the performance difference between a one-fits-all policy and the optimal policy in an average sense. Fig. \ref{plot}. demonstrates the evolution of this quantity during learning for three cases. Overall, despite that there are oscillations due to the randomness of meta-gradient estimators, the ratios become sufficiently small after adequate iterations, which implies the effectiveness of the algorithm.

\section{CONCLUSIONS}
In this paper, we studied the problem of meta-RL for a set of similar LQRs. In analogy to MAML, we formulated the objective \eqref{obj}, the goal is to refine a policy that attains good performance for the set of LQR problems utilizing direct gradient methods. 
Our proposed method overcomes the parameter unknownness and function inexplicitness through Monte-Carlo sampling and zeroth-order technique, without estimating the policy Hessian, which is common in existing literature.
We defined meta-learnability in making assumption \ref{admissible} and analyzed the boundedness and Lipschitz property of objective gradient, which immediately implies the convergence of exact gradient descent. 
A numerical experiment is given as an empirical assessment for the algorithm, which displays promising performance under the setting of the average cost difference ratio metric.
In future research, we plan to quantify the iteration and sample complexity bound of algorithm \ref{metalqr}, and gain more understanding for the geometry of objective function \eqref{obj}.




\section*{APPENDIX}

\proof[Proof of Prop. \ref{prop1}]{
 For arbitrary system $i$, consider a stable policy $K$ such that $\rho(A_i-B_iK) < 1$, define operator $\mathcal{T}_K(\Sigma)$ by:
 \begin{equation*}
      \mathcal{T}^i_K(\Sigma) = \sum_{t\geq 0} (A_i-B_i K)^t \Sigma [(A_i-B_i K)^t]^{\top} .
 \end{equation*}
 Here, $\mathcal{T}^i_K$ is an adjoint operator, observing that for any two symmetric positive definite matrices $\Sigma_1$ and $\Sigma_2$, we have
 \begin{equation*}
    \begin{aligned}
         \operatorname{Tr}( \Sigma_1 \mathcal{T}^i_K(\Sigma_2)) & =  \operatorname{Tr}(\sum_{t\geq 0} \Sigma_1 (A_i-B_i K)^t \Sigma_2 [(A_i-B_i K)^t]^{\top}) \\
         & = \operatorname{Tr} (\sum_{t\geq 0} [(A_i-B_i K)^t]^{\top} \Sigma_1 (A_i-B_i K)^t \Sigma_2 )\\ & = \operatorname{Tr}(\mathcal{T}^{i\top}_K(\Sigma_1) \Sigma_2)
    \end{aligned}
 \end{equation*}
 Meanwhile, since we know that $\Sigma_K^i$ satisfies recursion \eqref{xgramian},
$\Sigma^i_K = \mathcal{T}_K^i(\Psi_i) .$
Thus the average cost of $K$ for system $i$ can be written as
\begin{equation*}
\begin{aligned}
         J_i(K) &=  \operatorname{Tr}\left[\left(Q_i+K^{\top} R_i K\right) \cdot \Sigma^i_{K}\right] \\ &=\operatorname{Tr}\left[\left(Q_i + K^{\top} R_i K\right) \cdot \mathcal{T}_{K}^i\left(\Psi_i \right)\right]  \\ & =\operatorname{Tr}\left[\mathcal{T}_{K}^{i\top}\left(Q_i+K^{\top} R_i K\right) \cdot \Psi_i \right]=\operatorname{tr}\left(P^i_{K} \Psi_i\right).
\end{aligned}
\end{equation*}
By rule of product:
\begin{equation*}
\begin{aligned}
  \nabla J_i(K) & = 2R_i K\Sigma_K^i + \nabla \operatorname{Tr}(Q_i^{\prime} \mathcal{T}_K^i(\Psi_i))|_{Q^{\prime} = Q_i + K^{\top}R_i K} 
\end{aligned}
\end{equation*}
Here, we derive the expression for the second term. 
For symmetric positive definite matrix $\Sigma$, define operator $\Gamma^i_K(\Sigma) := (A_i - B_i K) \Sigma (A_i - B_i K)^{\top}$, we have 
\begin{equation*}
  Q^{\prime}_i \mathcal{T}^i_K (\Sigma^i_K) = Q^{\prime}_i\Psi_i +  \Gamma^i_K ( \mathcal{T}_K^i(\Sigma_K^i)),
\end{equation*}
and 
$\mathcal{T}^i_K  (\Sigma)= \sum_{t = 0}^{\infty} (\Gamma^i_K)^t(\Sigma).$
Since $\mathcal{T}_K^i$ is linear and adjoint
$$
 \operatorname{Tr}(Q_i^{\prime} \mathcal{T}_K^i(\Psi_i)) = \operatorname{Tr}(Q_i^{\prime} \Psi_i) + \operatorname{Tr}( \Gamma^{i\top}_K(Q^{\prime}_i)\mathcal{T}_K^{i\top}(\Psi_i)).
$$
Take derivative on both sides and unfold the right-hand side:
\begin{equation*}
\begin{aligned}
    \nabla \operatorname{Tr}(Q_i^{\prime} \mathcal{T}_K^i &(\Psi_i))  = \nabla \operatorname{Tr}(Q_i^{\prime} \Psi_i ) + \nabla \operatorname{Tr} (\Gamma^{i \top}_K(Q^{\prime}_i) )  \\ & \quad  \quad + \nabla \operatorname{Tr}(Q^{\prime \prime}_i \mathcal{T}_K^i(\Psi_i)) |_{Q^{\prime\prime} = \Gamma_K^{i\top}(Q^{\prime}_i)} \\
    & = -2B_i^{\top}[\sum_{t = 0}^{\infty} (\Gamma^{i,\top}_K)^t(Q^{\prime}_i)](A_i -B_i K)\mathcal{T}_K^i(\Psi_i) \\
    & = -2B_i^{\top} \mathcal{T}_K^{i, \top}(Q_i + K^{\top} R_i K) (A_i - B_i K) \Sigma^i_K, 
\end{aligned}
\end{equation*}
where we leverage the condition that spectrum $\rho(A_i - B_i K) < 1$, by which we have:
\begin{equation*}
    \operatorname{Tr}((\Gamma^{i,\top}_K)^t  Q^{\prime}_i) \leq \| Q_i^{\prime} \| \| A_i - B_i K\|^{2 t}  \underset{t \to \infty}{ \rightarrow }0, 
\end{equation*}
thus the series converge.
Combining with the fact that $P_K^i$ is actually the solution to the fixed point equation:
$   P^i_K = \mathcal{T}_K^i (Q_i + K^{\top} R_i K)$,
we get the desired result. 
\begin{equation*}
\begin{aligned}
   \nabla J_i(K) & = 2\left[\left(R_i+B_i^{\top} P_{K} B_i\right) K-B_i^{\top} P^i_{K} A_i\right] \Sigma^i_{K} 
\end{aligned}
\end{equation*}
\qedhere
}




\bibliographystyle{abbrv}
\bibliography{ref}

\begin{thebibliography}{10}

\bibitem{abbasi2011online}
Y.~Abbasi-Yadkori, D.~P{\'a}l, and C.~Szepesv{\'a}ri.
\newblock Online least squares estimation with self-normalized processes: An
  application to bandit problems.
\newblock {\em arXiv preprint arXiv:1102.2670}, 2011.

\bibitem{DBLP:journals/corr/Abbasi-YadkoriS14}
Y.~Abbasi{-}Yadkori and C.~Szepesv{\'{a}}ri.
\newblock Bayesian optimal control of smoothly parameterized systems: The lazy
  posterior sampling algorithm.
\newblock {\em CoRR}, abs/1406.3926, 2014.

\bibitem{pmlr-v54-abeille17b}
M.~Abeille and A.~Lazaric.
\newblock {Thompson Sampling for Linear-Quadratic Control Problems}.
\newblock In A.~Singh and J.~Zhu, editors, {\em Proceedings of the 20th
  International Conference on Artificial Intelligence and Statistics},
  volume~54 of {\em Proceedings of Machine Learning Research}, pages
  1246--1254, Fort Lauderdale, FL, USA, 20--22 Apr 2017. PMLR.

\bibitem{bradtke1994adaptive}
S.~J. Bradtke, B.~E. Ydstie, and A.~G. Barto.
\newblock Adaptive linear quadratic control using policy iteration.
\newblock In {\em Proceedings of 1994 American Control Conference-ACC'94},
  volume~3, pages 3475--3479. IEEE, 1994.

\bibitem{pmlr-v97-cohen19b}
A.~Cohen, T.~Koren, and Y.~Mansour.
\newblock Learning linear-quadratic regulators efficiently with only $\sqrt{T}$
  regret.
\newblock In K.~Chaudhuri and R.~Salakhutdinov, editors, {\em Proceedings of
  the 36th International Conference on Machine Learning}, volume~97 of {\em
  Proceedings of Machine Learning Research}, pages 1300--1309. PMLR, 09--15 Jun
  2019.

\bibitem{fallah2020provably}
A.~Fallah, K.~Georgiev, A.~Mokhtari, and A.~Ozdaglar.
\newblock Provably convergent policy gradient methods for model-agnostic
  meta-reinforcement learning.
\newblock {\em arXiv preprint arXiv:2002.05135}, 2020.

\bibitem{fazel2018global}
M.~Fazel, R.~Ge, S.~Kakade, and M.~Mesbahi.
\newblock Global convergence of policy gradient methods for the linear
  quadratic regulator.
\newblock In {\em International Conference on Machine Learning}, pages
  1467--1476. PMLR, 2018.

\bibitem{fiechter1997pac}
C.-N. Fiechter.
\newblock Pac adaptive control of linear systems.
\newblock In {\em Proceedings of the tenth annual conference on Computational
  learning theory}, pages 72--80, 1997.

\bibitem{finn2017model}
C.~Finn, P.~Abbeel, and S.~Levine.
\newblock Model-agnostic meta-learning for fast adaptation of deep networks.
\newblock In {\em International Conference on Machine Learning}, pages
  1126--1135. PMLR, 2017.

\bibitem{gershwin1970discrete}
S.~B. Gershwin and D.~H. JACOBSON.
\newblock A discrete-time differential dynamic programming algorithm with
  application to optimal orbit transfer.
\newblock {\em AIAA Journal}, 8(9):1616--1626, 1970.

\bibitem{DBLP:journals/corr/abs-2011-10300}
B.~Hambly, R.~Xu, and H.~Yang.
\newblock Policy gradient methods for the noisy linear quadratic regulator over
  a finite horizon.
\newblock {\em CoRR}, abs/2011.10300, 2020.

\bibitem{molybog2020global}
I.~Molybog and J.~Lavaei.
\newblock Global convergence of maml for lqr.
\newblock {\em arXiv preprint arXiv:2006.00453}, 2020.

\bibitem{williams1992simple}
R.~J. Williams.
\newblock Simple statistical gradient-following algorithms for connectionist
  reinforcement learning.
\newblock {\em Machine learning}, 8(3):229--256, 1992.

\bibitem{DBLP:journals/corr/abs-1907-06246}
Z.~Yang, Y.~Chen, M.~Hong, and Z.~Wang.
\newblock On the global convergence of actor-critic: {A} case for linear
  quadratic regulator with ergodic cost.
\newblock {\em CoRR}, abs/1907.06246, 2019.

\end{thebibliography}

\end{document}